\documentclass[twocolumn]{aastex631}

\usepackage{multirow}
\usepackage{orcidlink}
\usepackage{threeparttable}
\usepackage{tabularx}
\usepackage{CJKutf8}
\usepackage{graphicx}
\usepackage{amsmath}
\usepackage{ulem}
\usepackage{soul}

\begin{document}
\begin{CJK}{UTF8}{gbsn}
\title{The Study of Mode Switching behavior of PSR J0614+2229 Using the Parkes Ultra-wideband Receiver Observations }

\author{Yanqing Cai}
\affiliation{School of Physics and Electronic Science, Guizhou Normal University, Guiyang 550025, China}

\author[0000-0002-2060-5539]{Shijun Dang}
\email{dangsj@gznu.edu.cn(S.J.Dang)}
\affiliation{School of Physics and Electronic Science, Guizhou Normal University, Guiyang 550025, China}
\affiliation{Guizhou Radio Astronomical Observatory, Guizhou University, Guiyang 550025, China}
\affiliation{Guizhou Provincial Key Laboratory of Radio Astronomy and Data Processing, Guiyang 550025, China}

\author{Rai Yuen}
\affiliation{XinJiang Astronomical Observatory, CAS, Urumqi, Xinjiang 830011, People's Republic of China}

\author{Lunhua Shang}
\email{shanglh@gznu.edu.cn(L.H.Shang)}
\affiliation{School of Physics and Electronic Science, Guizhou Normal University, Guiyang 550025, China}

\author[0000-0002-9786-8548]{Feifei Kou}
\affiliation{XinJiang Astronomical Observatory, CAS, Urumqi, XinJiang 830011, People's Republic of China}

\author{Jianping Yuan}
\affiliation{XinJiang Astronomical Observatory, CAS, Urumqi, XinJiang 830011, People's Republic of China}

\author[0000-0001-8539-4237]{Lei Zhang}
\affiliation{National Astronomical Observatories, Chinese Academy of Sciences, Beijing 100101, China.\\}

\author{Zurong Zhou}
\affiliation{National Time Service Center, Chinese Academy of Sciences, Xi’an 710600, China.\\}
\affiliation{Key Laboratory of Time Reference and Applications, Chinese Academy of Sciences, Xi’an 710600, China\\}

\author[0000-0003-2991-7421]{Na Wang}
\affiliation{XinJiang Astronomical Observatory, CAS, Urumqi, XinJiang 830011, People's Republic of China}

\author{Qingying Li}
\affiliation{School of Physics and Electronic Science, Guizhou Normal University, Guiyang 550025, China}

\author[0000-0003-4705-5255]{Zhigang Wen}
\affiliation{XinJiang Astronomical Observatory, CAS, Urumqi, XinJiang 830011, People's Republic of China}

\author[0000-0002-7662-3875]{Wenming Yan}
\affiliation{XinJiang Astronomical Observatory, CAS, Urumqi, XinJiang 830011, People's Republic of China}

\author[0000-0003-4498-6070]{Shuangqiang Wang}
\affiliation{XinJiang Astronomical Observatory, CAS, Urumqi, XinJiang 830011, People's Republic of China}

\author{Shengnan Sun}
\affiliation{XinJiang Astronomical Observatory, CAS, Urumqi, XinJiang 830011, People's Republic of China}

\author{Habtamu Menberu Tedila}
\affiliation{National Astronomical Observatories, Chinese Academy of Sciences, Beijing 100101, China.\\}

\author{Shuo Xiao}
\affiliation{School of Physics and Electronic Science, Guizhou Normal University, Guiyang 550025, China}
\affiliation{Guizhou Provincial Key Laboratory of Radio Astronomy and Data Processing, Guiyang 550025, China}

\author{Xin Xu}
\affiliation{School of Physics and Electronic Science, Guizhou Normal University, Guiyang 550025, China}

\author{Rushuang Zhao}
\affiliation{School of Physics and Electronic Science, Guizhou Normal University, Guiyang 550025, China}

\author[0000-0001-9389-5197]{Qijun Zhi}
\affiliation{School of Physics and Electronic Science, Guizhou Normal University, Guiyang 550025, China}

\author{Aijun Dong}
\affiliation{School of Physics and Electronic Science, Guizhou Normal University, Guiyang 550025, China}

\author{Bing Zhang}
\affiliation{School of Physics and Electronic Science, Guizhou Normal University, Guiyang 550025, China}

\author{Wei Li}
\affiliation{School of Physics and Electronic Science, Guizhou Normal University, Guiyang 550025, China}

\author{Yingying Ren}
\affiliation{School of Physics and Electronic Science, Guizhou Normal University, Guiyang 550025, China}


\author{Yujia Liu}
\affiliation{School of Physics and Electronic Science, Guizhou Normal University, Guiyang 550025, China}


\begin{abstract}
In this paper, we presented a detailed single pulse and polarization study of PSR J0614+2229 based on the archived data observed on 2019 August 15 (MJD 58710) and September 12 (MJD 58738) using the Ultra-wideband Low-frequency Receiver on the Parkes radio telescope. The single-pulse sequences show that this pulsar switches between two emission states, in which the emission of state A occurs earlier than that of state B in pulse longitude. We found that the variation in relative brightness between the two states is related to time and both states follow a simple power law very well. Based on the phase-aligned multi-frequency profiles, we found that there is a significant difference in the distributions of spectral index across the emission regions of the two states. 
Furthermore, we obtained the emission height evolution for the two emission states and found that, at a fixed frequency, the emission height of state A is higher than that of state B. What is even more interesting is that the emission heights of both states A and B have not changed with frequency.  Our results suggest that the mode switching of this pulsar is possibly caused by changes in the emission heights that alter the distributions of spectral index across the emission regions of states A and B resulting in the frequency-dependent behaviors, i.e., intensity and pulse width. 
\end{abstract}

\keywords{pulsars: general -- pulsars: individual: PSR J0614+2229}


\section{Introduction} \label{sec:intro}
Pulsars,  which can emit electromagnetic waves, are quickly rotational neutron stars and they are recognized as the most stable rotators in the universe because of their rotational clock-like stability and their highly repeatable pulse shapes. To date, a rich variety of time-dependent phenomena has long been observed in radio pulsars. A famous example of such a phenomenon is mode switching, namely the pulse profiles of some pulsars are observed to switch between two or more shape states, with a time-scale ranging from several to thousands of periods \citep{1970Natur.228.1297B,2007MNRAS.377.1383W}. A recent addition to these phenomena is ‘emission shifts’ (also known as 'swooshes') where the regular pulsed emission exhibited a progressive shift towards earlier longitudes over a few pulses and then back to its usual longitudinal location after remaining in early for typically several tens of pulses \citep{2006MNRAS.370..673R}. The origin of mode switching and swooshes has remained a mystery since the discovery of them. There is increasing evidence that the mode changing or swooshes is possibly caused by changes of magnetospheric particle current ﬂow (\citeauthor{2010Sci...329..408L} \citeyear{2010Sci...329..408L}; \citeauthor{2017MNRAS.469.2049Y} \citeyear{2017MNRAS.469.2049Y}; \citeauthor{2021MNRAS.506.5836R} \citeyear{2021MNRAS.506.5836R}). However, the underlying physical mechanism for the change in magnetosphere state is not clear yet.




PSR J0614$+$2229 (B0611$+$22) is a very young and peculiar pulsar, which was discovered by using the Jodrell Bank Mark IA radio telescope (\citeauthor{1972Natur.240..229D} \citeyear{1972Natur.240..229D}) and was confirmed as a mode switching pulsar with two different emission states (\citeauthor{hankins1992magnetospheric} \citeyear{hankins1992magnetospheric}). Since the discovery of PSR J0614$+$2229, emission behavior about it at multiple frequencies has been well studied. \citeauthor{2014MNRAS.439.3951S} (\citeyear{2014MNRAS.439.3951S}) first reported this pulsar has a burst emission phenomenon and the two emission states offset in the pulse phase at both 327 MHz and 1400 MHz. Subsequently, observations observed simultaneously at 150/327 MHz and 150/820 MHz revealed that the emission state with earlier longitude is stronger at 150 MHz and 327 MHz than other states, but weaker at 820 MHz. \citeauthor{2016MNRAS.462.2518R} (\citeyear{2016MNRAS.462.2518R}) speculated that this phenomenon might be related to the different spectral indices of the two states. This conjecture was confirmed by \citeauthor{2020ApJ...890...31Z} (\citeyear{2020ApJ...890...31Z}). They found that the two emission states have different spectrum indexes and suggested that the differences in the frequency-dependent behaviors of intensity and pulse width might be caused by two opposite types of spectral index variation across the beam for the two states. Moreover, based on observation using the Five-hundred-meter Aperture Spherical radio Telescope, \citeauthor{2022ApJ...934...57S} (\citeyear{2022ApJ...934...57S}) suggested that the mode switching of this pulsar is related to the changes in the emission heights. Nevertheless, it is still unknown for the emission mechanism of this pulsar. In this paper, we carried out a detailed single pulse and polarization study of PSR J0614+2229 based on the data observed with the Parkes UWL receiver, which may provide further insights into the emission characteristics of this pulsar. In Section \ref{sectwo}, we give detailed information about our observations using the Parkes 64m radio telescope. Our results are presented in Section \ref{secthree}. And our discussions based on results are in Section \ref{secfour}. Conclusions are given in Section \ref{secfive}.

\section{Observations and Data Processing} \label{sectwo}
We collected two available UWL data observed by the Parkes radio telescope on 2019 August 15 (MJD 58710) and September 12 (MJD 58738). The bandwidth of the UWL receiver is from 704 to 4032 MHz and is subdivided into 3328 frequency channels. The time resolution is 64 $\mu s$. The detailed information about the UWL receiver and backends are described in \citeauthor{2013PASA...30...17M} (\citeyear{2013PASA...30...17M}) and \citeauthor{2020PASA...37...12H} (\citeyear{2020PASA...37...12H}).

For offline data reduction, firstly, we used the \textsc{DSPSR} software package (\citeauthor{2011PASA...28....1V} \citeyear{2011PASA...28....1V}) to extract individual pulses from raw data and remove inter-channel dispersion delays. Then, we used the program \textsc{PAZ} in the \textsc{PSRCHIVE} software package (\citeauthor{2014PASA...31...41H} \citeyear{2014PASA...31...41H}) to eliminate the RFI automatically. To further optimize the data, we utilized the \textsc{PSRCHIVE} program \textsc{PAZI} to remove frequency channels or sub-integrations influenced by RFI manually. Subsequently, we calibrated both the polarization and flux of the data using the \textsc{PAC} plugin of the \textsc{PSRCHIVE} software package. Finally, for better analysis of multi-band emission characteristics, the data from both observations were divided into eight separated subbands and the bandwidth of high-frequency is relatively wide due to the weak signal-to-noise ratio. The summaries of the observations are described in Table \ref{table1}, including Modified Julian Date(MJD), the centered frequency($\nu$), bandwidth($\Delta$$\nu$), the specified duration for every sub-integration($T_{s u b}$), total quantity of sub-integrations($N_{s u b}$) and number of bins in each average profile($N_{b i n}$). 
\begin{table}[h!]
    \centering
    \caption{Summary of Parkes Observations}
    \label{table1}
     
\begin{threeparttable}
\begin{tabular}{cccccc}
     \hline
     \hline
     MJD & $\nu$ & $\Delta\nu$ & $T_{sub}$ & $N_{sub}$ & $N_{bin}$  \\
     (d) &(MHz) & (MHz) & (s) & \\
     \hline
     \multirow{8}{*}{58710} 
      & 879  & 350 & 15 & 460 & 1024\\
     
     & 1229 & 350 & 15 & 460 &1024 \\
     
      &  1579 & 350 & 15 & 460 & 1024\\
     
     & 1972 &350  & 15 &  460& 1024\\
     
   & 2279 & 350 & 15 & 460 & 1024 \\
     
     & 2629 & 350 & 15 &  460& 1024\\
     
      & 3111 & 614 & 15& 460& 1024\\
     
     & 3725 & 614 &15  & 460 & 1024\\
     \hline
\end{tabular}
\begin{tablenotes}
    \footnotesize
    \item Note. the values of $\nu$, $\Delta$$\nu$, $T_{s u b}$ and $N_{bin}$ are the same for the two observations. What's different is that the number of sub-integrations ($T_{s u b}$) at eight frequencies in the second observation(MJD 58738) is 286.    
    \end{tablenotes}
    \end{threeparttable}
\end{table}

\section{results} \label{secthree}
\begin{figure}
    \includegraphics[width=\columnwidth]{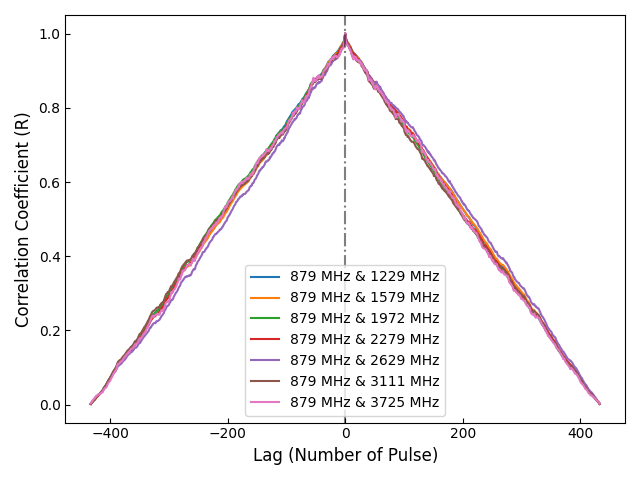}
    \caption{The cross-correlations between the pulse energies for the sequence of sub-integrations throughout the on$-$pulse window at a center frequency of 879 MHz and other frequencies in the first observation.}
    \label{one}
\end{figure}

\begin{figure}[h!]
    \includegraphics[width=\columnwidth]{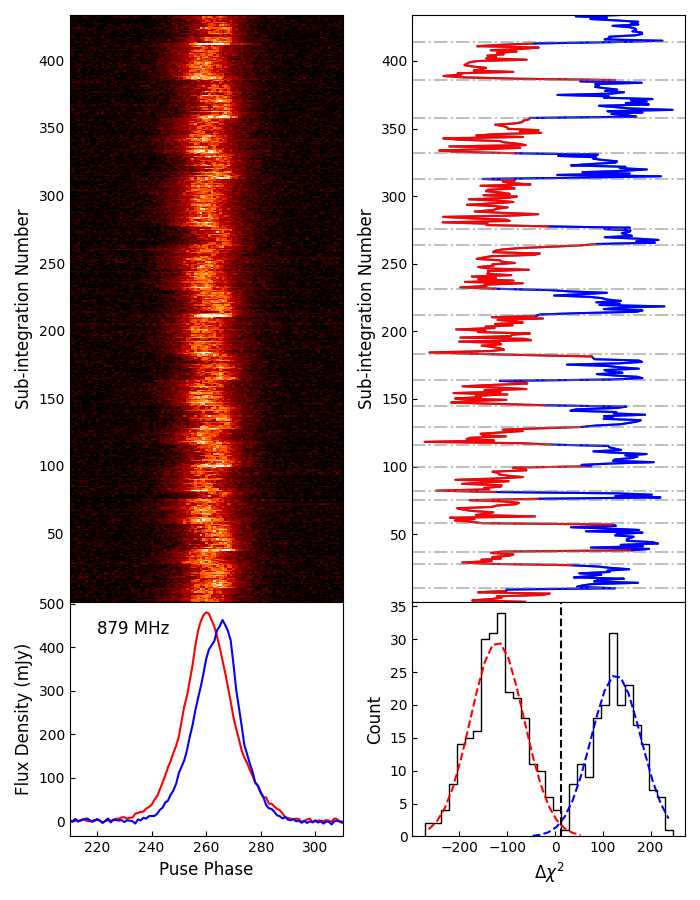}
    \caption{The first observation (MJD 58710) with the center frequency of 879 MHz. Top left panel: the sequence of sub-integrations from PSR J0614+2229. Top right panel: the $\Delta$$\chi_i$ for the corresponding sub-integration on the left, where the red and blue sequences represent states A and B, respectively. Bottom left panel: the mean pulse profiles for state A (red line) and state B (blue line). Bottom right panel: histogram of the $\Delta$$\chi_i$ for the entire sub integration is shown with the black solid line. In which, the vertical black dashed line represents the threshold value based on the combination of two Gaussian components (red and blue dotted lines) that distinguishes the two states. }
    \label{two}
\end{figure}

\begin{figure}[h!]
    \includegraphics[width=\columnwidth]{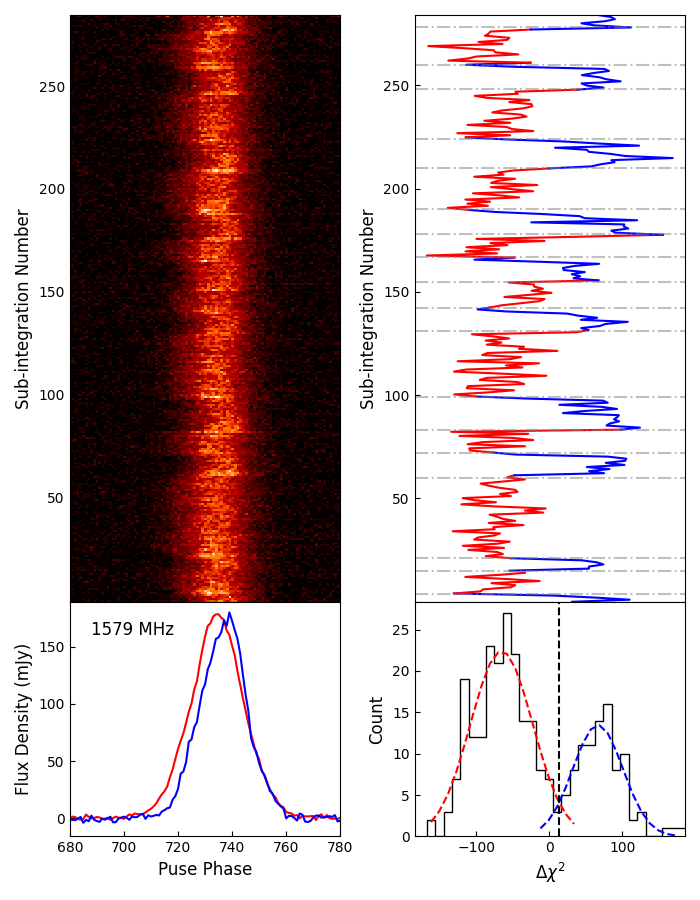}
    \caption{Same as Figure \ref{two} but for the second observation (MJD 58738) with the center frequency of 1579 MHz. }
    \label{three}
\end{figure}

\subsection{Two emission states} \label{three_sub_one}

To check whether the state switching occurs simultaneously for each frequency, we performed a cross-correlation analysis between the pulse energy of the sub-integral sequence within the pulse window for each sub-bandwidth and the pulse energy of the sub-integral sequence with a center frequency of 879 MHz in the first observation. The result is shown in Figure \ref{one}, it can be seen that the cross-correlation reaches its maximum value of 1 at zero lag. This means that states A and B change simultaneously at different frequencies.

To determine the mode-changing properties, we followed the method described in \citet{2018ApJ...867L...2M} and defined a quantitative metric for classifying individual pulses as belonging to states A or B:
\begin{equation*}
    \Delta\chi_i^2 = \sum_{\phi}\frac{(p_i(\phi) - p_a(\phi))^2 - (p_i(\phi)-p_b (\phi))^2}{\sigma(\phi)^2},
\end{equation*}
where, $\phi$ is the phase bin, $p_i(\phi)$ is an individual pulse profile, $p_a(\phi)$ and $p_b(\phi)$ are the average pulse profiles for two states, and $\sigma(\phi)$ is the standard deviation from the mean of the entire data set. For the first observation (MJD 58710), we chose a subband with a center frequency of 879 MHz as the distinguishing template because it has a high signal-to-noise ratio. The mode metric $\Delta$$\chi_i$ corresponding to individual pulses for this template are shown in the bottom right panel of Figrue \ref{two}, which displays two Gaussian-like components. We used Gaussian functions to ﬁt these two components and selected the intersection of the two ﬁtted Gaussian profiles as the threshold value (the vertical dashed line in Figure \ref{two}). And then, if the $\Delta$$\chi_i$ of a subintegration is below this value, we classify it as being in state A; otherwise, it belongs to state B. In this way, we can obtain the $\Delta$$\chi_i$ sequences corresponding to states A and B separately, and label the sequences corresponding to states A and B in red and blue, respectively. The marked $\Delta$$\chi_i$ sequences are shown in the top-right panel of Figure \ref{two}, which displays a clear correspondence with the sequence of subintegrations in the top-left panel of this figure. The average pulse profiles for state A (red solid line) and state B (blue solid line) are shown in the bottom-left panel of Figure \ref{two}, where the proﬁle of state A is slightly earlier than that of state B in longitude. For states A and B on other frequencies, we can distinguish them based on the marked $\Delta$$\chi_i$ sequences obtained at the center frequency on 879 MHz above. And for the second observation (MJD 58738), the procedure to identify the two states is the same as the first observation, and what's different is that the subband we select as the distinguishing template is one with a center frequency of 1579 MHz. The result is shown in Figure \ref{three}. For the first observation, this pulsar spends 57\% of the time in state A and 43\% of the time in state B. For the second observation, the proportions of state A and state B are 66\% and 34\%, respectively.

\begin{figure*}[th!]
   \includegraphics[width=7in,height=9in]{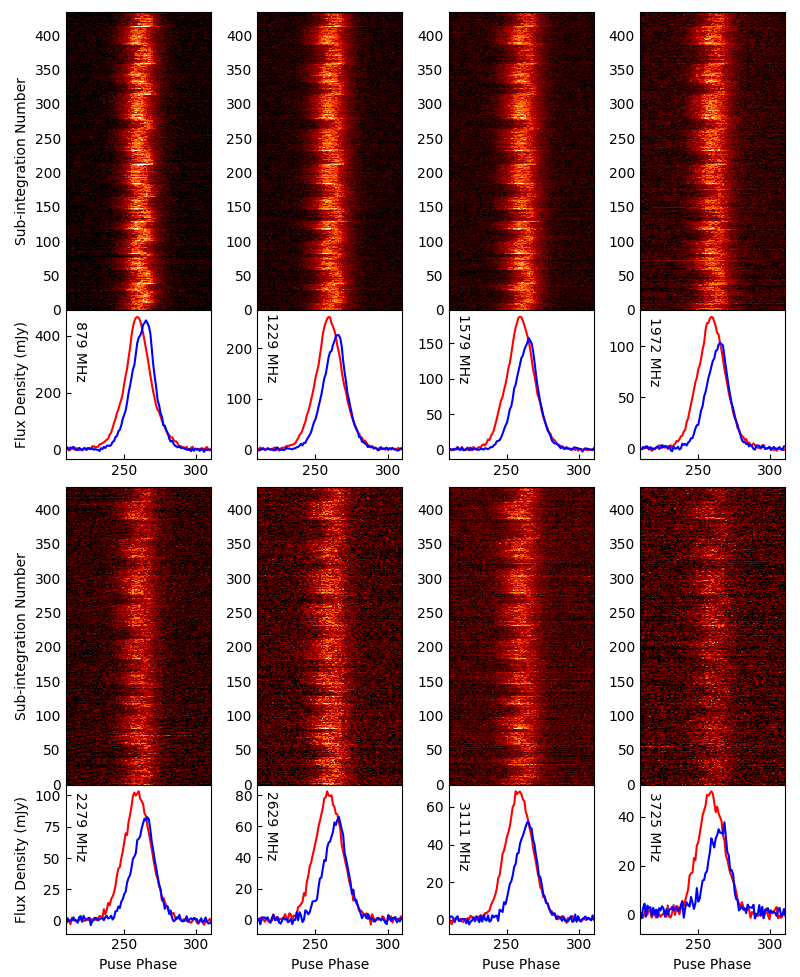}
   \caption{The result of the first observation (MJD 58710). Plots displaying the sequence of subintegrations for PSR J0614+2229 at eight different frequencies across UWL band, with average pulse profiles for state A (red) and state B (blue).}
   \label{cyq4}
\end{figure*}
\begin{figure*}[th!]
   \includegraphics[width=7in,height=9in]{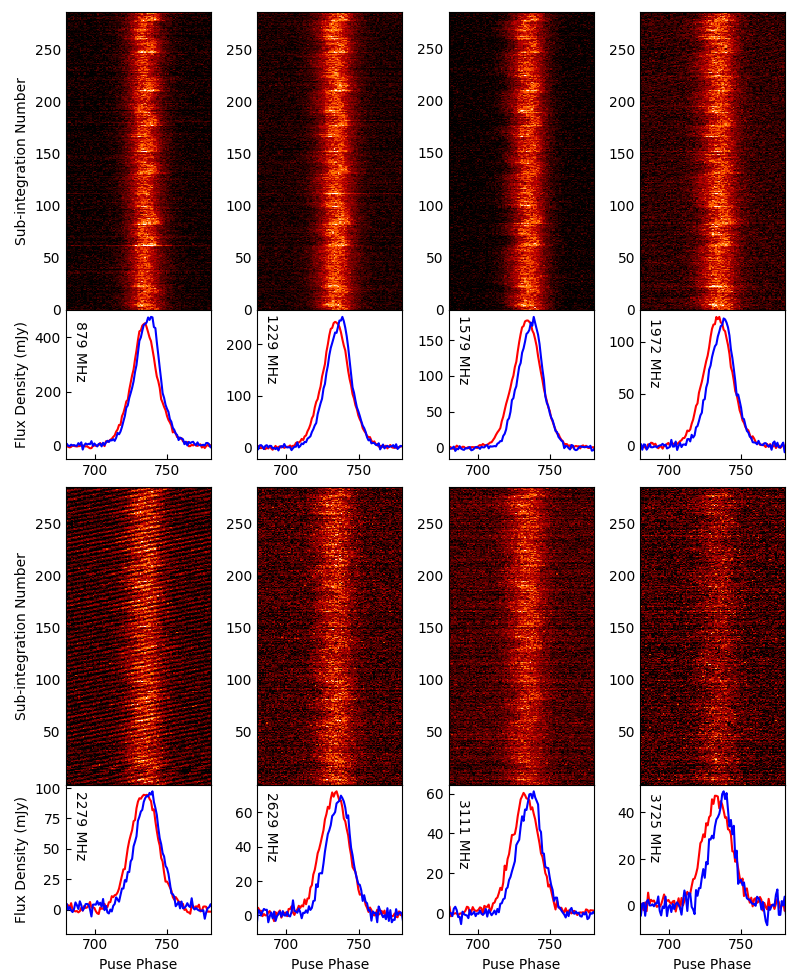}
   \caption{Same as Figure \ref{cyq4} but for the second observation (MJD 58738). Note that the radio frequency interference (RFI) at higher frequencies is relatively stronger, especially at 2279 MHz.}
   \label{cyq5}
\end{figure*}

\begin{figure}[h!]
    \includegraphics[width=\columnwidth]{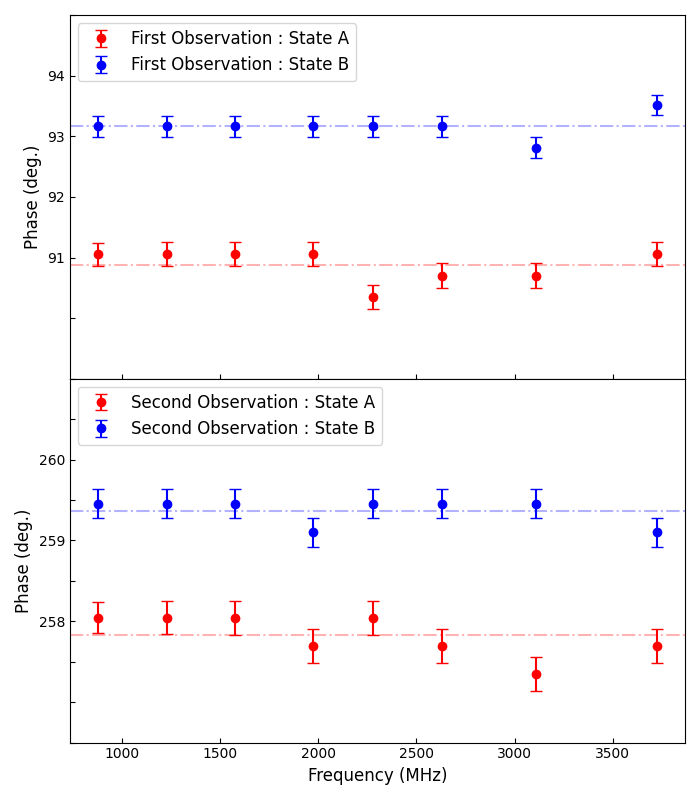}
    \caption{Top: the phase of pulse peaks for states A (red dot) and B (blue dot) in the first observation versus frequency. Bottom: same as the top panel but for the second observation. The horizontal dash-dotted lines represent the mean values of the phase of pulse peaks for the two emission states at a wide frequency range.}
    \label{cyq6}
\end{figure} 
The color-coded pulse sequence diagrams of the two observations are shown in the top panels of Figure \ref{cyq4} and Figure \ref{cyq5}, 
an intriguing phenomenon from the two figures is that the direction of phase shift in the two observations at the center frequency of 879 MHz seems to be different, whereby in the first observation the phase shift occurs in state A while in the second observation it occurs in state B, this phenomenon is similar with \citet{2016MNRAS.462.2518R} but for the different frequencies in their paper (see Figure 2 of that paper).  

\subsection{Frequency dependence of phase offset and pulse width} \label{sec3}
\begin{figure}
    \centering
    \includegraphics[width=\columnwidth]{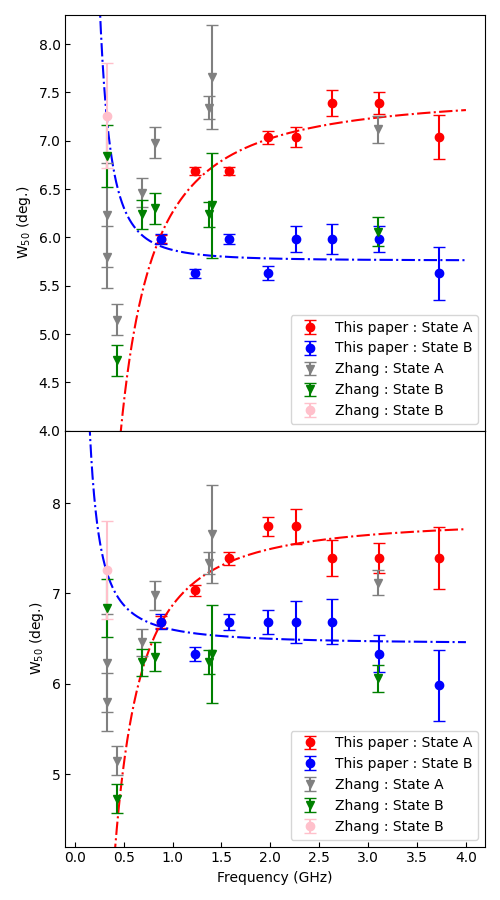}
    \caption{The top and bottom panels represent FWHM versus observing frequency for the first and the second observations, respectively. Red and blue symbols correspond to state A and state B. The best-fit Thorsett relationship was plotted as the red and blue dashed-dotted curves. The meaning of pink, green, and gray symbols can be seen text for details.}
    \label{cyq7}
\end{figure}
The average pulse profiles for state A (red solid line) and state B (blue solid line) at different frequencies are shown in the bottom panels of Figure \ref{cyq4} and Figure \ref{cyq5}. To investigate the frequency-dependent evolution of the phase shift in the two emission states, we utilized Gaussian fitting to determine the phase difference at the peak of the integrated profile for states A and B at each frequency (see Figure~\ref{cyq6}). 
It is noteworthy that the peak phase shift of the two emission states does not change significantly with frequency. This indicates that the phase of the two emission states is relatively stable over a wide frequency range.

In addition, studying the evolution of the profile width with observing frequency can help us better understand the structure of the radiation region. The profile width at 50\% ($W_{50}$) of the peak amplitude is calculated for the eight frequency bands in the two observations and the uncertainty of $W_{50}$ is computed as follows: \begin{equation}
   \sigma = res \sqrt{1+(\frac{rms}{I})^2}, 
\end{equation} where $res$ refers to the sampling time of per pulse bin, $rms$ is the standard deviation of the off-pulse-noise for the total profile, and Stokes I is measured signal intensity at the bin of interest. To investigate the trend of the frequency evolution of the profile width for the two emission states, the well-known Thorsett relationship $W_{50}$ = Y$\nu^{\mu}$ + R (\citeauthor{1991ApJ...377..263T} \citeyear{1991ApJ...377..263T}; \citeauthor{2014ApJS..215...11C} \citeyear{2014ApJS..215...11C}; \citeauthor{2022ApJ...934...57S} \citeyear{2022ApJ...934...57S}) was employed to fit the data. We performed Markov Chain Monte Carlo ﬁtting \citep{2019MNRAS.490.4565J} using the python package \textsc{emcee} (\citeauthor{2013PASP..125..306F} \citeyear{2013PASP..125..306F}) to determine the values for Y, $\mu$ and R, and these values are shown in Table \ref{table2}. Figure \ref{cyq7} shows the frequency evolution of the profile width for state A (red symbols) and state B (blue symbols), the top and bottom panels of this figure correspond to the observations at MJD 58710 and 58738, respectively. The symbols marked in pink, gray, and green are from \citep{2020ApJ...890...31Z}, who measured the $W_{50}$ of the two states based on previously published papers \citep[such as,][]{1980ApJ...239..310F,2014MNRAS.439.3951S,2016MNRAS.462.2518R} and archived data (See Section 3.2 and Table 3 of \citep{2020ApJ...890...31Z} for details). Here, the gray inverse-triangle represents the $W_{50}$ of state A, and the green inverse-triangles and pink dot correspond to state B. Note that we just employed the gray and green inverse-triangles to check the tendency of the $W_{50}$ versus frequency. However, the change of $W_{50}$ of state B with frequencies is relatively stable compared with state A. Therefore, value of $W_{50}$ at 327 MHz from \citeauthor{2020ApJ...890...31Z} (\citeyear{2020ApJ...890...31Z}) which marked in pink dot is selected to fit the evolution. The best-fit Thorsett relationship was plotted as the red and blue dashed-dotted curves in Figure \ref{cyq7}. We can see that the frequency dependence of the $W_{50}$ of the two states are opposite each other. The $W_{50}$ of state A increases with frequency, while that of state B decreases with frequency. This phenomenon is consistent with the result of \citet{2020ApJ...890...31Z}. Usually, the decreasing pulse width with increasing frequency is interpreted using a scenario of narrowband emission known as radius-to-frequency mapping \citep[RFM;][]{1970Natur.225..612K,1978ApJ...222.1006C}. However, the $W_{50}$  of state A increases with frequency, which does not align with this RFM model.

\begin{table}[h!]
    \centering
    \caption{List of the full width of half maximum fitting parameters for states A and B.}
    \label{table2}
    \begin{tabular}{ccccc}
    \hline
    \hline
     &  & Y & $\mu$ & R \\
    \hline
    \multirow{2}{*}{MJD: 58710}
    & State A & $-1.24_{-0.31}^{+0.22}$ & $-1.38_{-0.44}^{+0.37}$ & $7.50_{-0.20}^{+0.30}$ \\  
   
    &State B & $+0.11_{-0.05}^{+0.07}$ & $-2.33_{-0.51}^{+0.40}$ & $5.76_{-0.05}^{+0.04}$  \\
    \hline
    \multirow{2}{*}{MJD: 58738}
    &State A & $-0.96_{-0.33}^{+0.21}$ & $-1.51_{-0.59}^{+0.52}$ & $7.83_{-0.19}^{+0.32}$  \\
    & State B & $+0.16_{-0.09}^{+0.22}$ & $-1.44_{-0.71}^{+0.58}$ & $6.44_{-0.18}^{+0.09}$   \\
    \hline
    \end{tabular}
\end{table}.

\begin{figure}
    \centering
    \includegraphics[width=\columnwidth]{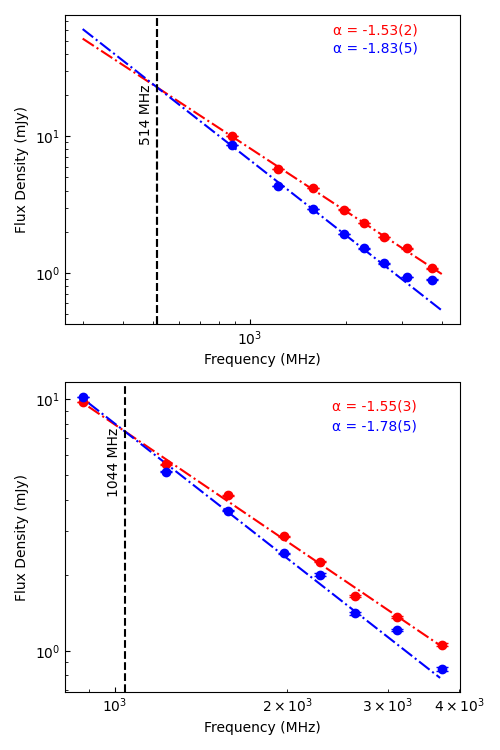}
    \caption{Top and bottom panels represent the results of the first and the second observations, respectively. Red and blue symbols correspond to the spectra of the states A and B. The red and blue dash-dotted lines show the best-fit power laws for our data. The vertical black line shows the intersection of the red and blue dash-dotted lines. The red $\alpha$ and the blue $\alpha$ in the upper right of each panel are the spectral indices for state A and state B, respectively.}
    \label{cyq8}
\end{figure}

\begin{figure}[h]
    \centering
    \includegraphics[width=\columnwidth]{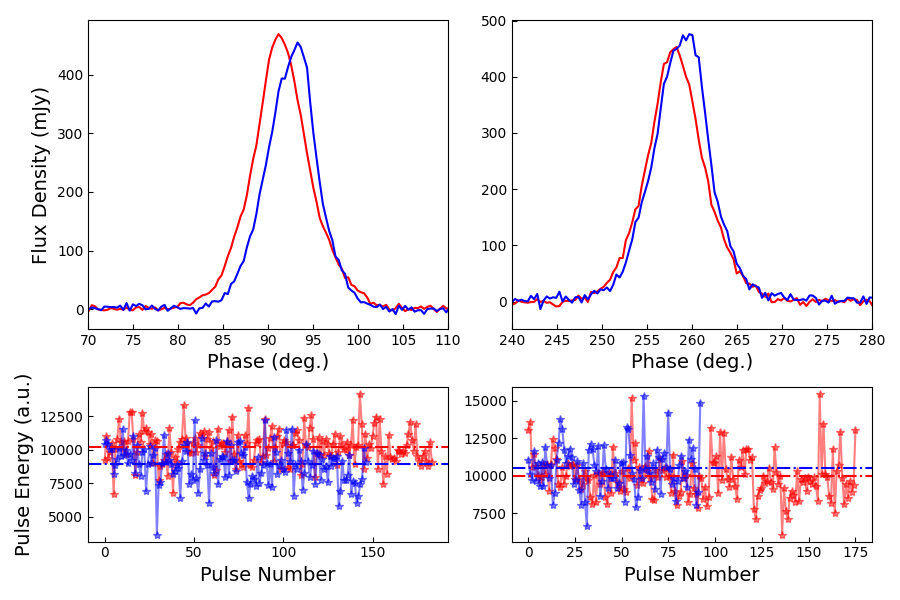}
    \caption{The left and right columns show the results of the first and second observations for the same frequency on 879 MHz, respectively. Top panels: integrated pulse profiles of state A (red lines) and state B (blue lines). Bottom panels: the pulse energy variations for the sequence of subintegrations throughout the on-pulse window, where the red and blue lines represent state A and state B, respectively. The horizontal dash-dotted lines represent the
    mean values of the pulse energy for different states. 
    }
    \label{cyq9}
\end{figure}

\subsection{The difference of spectral between the two states}
The observed flux densities of pulsars vary significantly over time because of diffractive interstellar scintillation (DISS) and refractive interstellar scintillation (RISS). To obtain reliable flux estimates, it is necessary to consider the biases given by DISS and RISS. DISS is due to constructive and destructive interference of radio waves emitted by pulsars during their propagation to the observer's location, with a time scale of approximately minutes \citep{2011A&A...534A..66L}. RISS occurs because of focusing and defocussing of the pulsar emission by the scattering medium and has a time-scale of days to months \citep{1994MNRAS.269.1035G}. In this paper, the observation length at MJD 58710 and 58738 were carried out approximately 2 hours and 72 minutes, respectively. This indicates that the ﬂux may not be dramatically modulated by the RISS and DISS. In addition, PSR J0614$+$2229 has a large dispersion measure (DM) value with 96 pc cm$^{-3}$. \citet{2000ApJ...539..300S} show that high DM pulsars at large distances have nearly constant observed flux densities over years, indicating that the pulsar emission is stable when individual pulses are integrated for at least a few hours. 

The flux densities of PSR J0614+2229, $S$, and it's uncertainty, $e$, for each observation at different frequencies were measured with the \textsc{psrchive} program \textsc{psrflux} as:
\begin{equation}
S = (\sum_{i}^{n_{on}} {I_{i}}) / n_{tot},
\end{equation}
and 
\begin{equation}
e = \sigma \sqrt{(n_{on})}/n_{tot},
\end{equation}
where \textit{$I_i$}, $\sigma$, $n_{tot}$ and $n_{on}$ are the flux density at $i$-th phase bin, the root-mean-square of the 'off-pulse' profile, the total account of phase bins in each period and the 'on' part of the profile, respectively. The measured values are listed in Table \ref{tablethree}, where the F$\textit{I}_A$ and F$\textit{I}_B$ corresponds to flux densities for states A and B in the first observation (MJD 58710), S$\textit{I}_A$ and S$\textit{I}_B$ same as F$\textit{I}_A$ and F$\textit{I}_B$ but for the second observation (MJD 58738). The number of data points for states A and B is enough to perform a least-square fitting using a function of a simple power law \citep{2018MNRAS.473.4436J}. The fitted results are shown in Figure \ref{cyq8}, where both states A and B follow a simple power law very well. The best-fitting spectral indices for states A and B are $-$1.53(2) and $-$1.83(5) in the first observation, $-$1.55(3) and $-$1.78(5) in the second observation. In both observations, the spectrum of state A was flatter than that of state B, which is consistent with the results of \citet{2020ApJ...890...31Z}. \begin{table}[h]
    \caption{Flux densities of states A and B at eight frequencies in two observations. Here, $\nu$ is the center frequency. F$\textit{I}_A$ and F$\textit{I}_B$ correspond to flux densities for state A and state B in the first observation, S$\textit{I}_A$ and S$\textit{I}_B$ same as F$\textit{I}_A$ and F$\textit{I}_B$ but for the second observation.}
    \label{tablethree}
    \begin{threeparttable}
    \begin{tabular}{ccccc}
    \hline
    \hline
    $\nu$ & F$\textit{I}_A$ & F$\textit{I}_B$ &S$\textit{I}_A$ & S$\textit{I}_B$    \\
  (MHz) & (mJy) & (mJy) & (mJy) & (mJy) \\
    \hline
    879 & 10.03(2) & 8.61(2) & 9.77(3) & 10.23(4) \\
    1229 & 5.78(1) & 4.33(1) & 5.52(1) & 5.16(2) \\
    1579 & 4.21(1) & 2.93(1) & 4.15(1) & 3.61(1) \\
    1972 & 2.91(1) & 1.93(1) & 2.86(1) & 2.45(1) \\
    2279 & 2.31(1) & 1.51(1) & 2.26(2) & 2.01(1) \\
    2629& 1.84(1) & 1.18(1) & 1.66(1) & 1.41(1) \\
    3111 & 1.52(1) & 0.93(1) & 1.37(1) & 1.21(1) \\
    3725 & 1.09(1) & 0.89(2) & 1.06(2) & 0.85(2) \\
    \hline 
    \end{tabular}
    \end{threeparttable}
\end{table}
\begin{table*}[t]
\renewcommand{\arraystretch}{1.5}
    \centering
    \caption{List of the PA-fitting Parameters}
    \label{tablefour}
    \begin{tabular}{cccccccccc}
    \hline
    \hline
    MJD & $\nu$ & $\alpha_A$ & $\alpha_B$ & $\beta_A$ & $\beta_B$ & $\phi_{0,A}$ & $\phi_{0,B}$ & $\psi_{0,A}$ & $\psi_{0,B}$ \\
    (d) & (MHz) & ($^\circ$) & ($^\circ$) & ($^\circ$) & ($^\circ$) & ($^\circ$) & ($^\circ$) & ($^\circ$)& ($^\circ$)  \\
    \hline
    \multirow{8}{*}{58710}
    & 879 & $125.62_{-22.29}^{+18.48}$ & $86.78_{-16.61}^{+14.59}$ & $4.51_{-1.56}^{+1.10}$ & $6.25_{-0.31}^{+0.27}$ & $98.85_{-0.26}^{+0.17}$ & $99.27_{-0.24}^{+0.20}$ &$133.77_{-3.13}^{+2.31}$  & $128.97_{-2.60}^{+2.36}$ \\
    & 1229 & $121.93_{-19.14}^{+19.13}$ & $90.05_{-18.10}^{+17.88}$ & $4.64_{-1.33}^{+0.92}$ & $5.70_{-0.44}^{+0.3} $& $98.62_{-0.21}^{+0.16}$ & $99.14_{-0.22}^{+0.21}$ & $133.19_{-2.70}^{+2.23}$ & $130.15_{-2.69}^{+2.75}$ \\
    & 1579 & $113.18_{-17.64}^{+17.11}$ & $87.51_{-17.48}^{+16.98}$ & $4.70_{-0.92}^{+0.53}$ & $5.08_{-0.36}^{+0.26}$ & $98.01_{-0.18}^{+0.15}$ & $98.66_{-0.17}^{+0.17}$ & $131.84_{-2.38}^{+2.08}$ & $130.64_{-2.53}^{+2.53}$  \\
    & 1972 & $141.79_{-24.92}^{+17.73}$ & $70.55_{-19.69}^{+22.88}$ & $2.99_{-1.34}^{+1.52}$ & $5.16_{-0.71}^{+0.41}$ & $97.99_{-0.18}^{+0.10}$ & $98.48_{-0.20}^{+0.24}$ & $135.51_{-2.75}^{+1.53}$ & $127.99_{-2.59}^{+3.18}$ \\
    & 2279 & $104.47_{-26.25}^{+25.65}$ & $85.69_{-26.06}^{+27.77}$ & $5.34_{-1.16}^{+0.54}$ & $6.09_{-1.04}^{+0.77}$ & $97.24_{-0.31}^{+0.27}$ & $97.94_{-0.39}^{+0.40}$ & $126.38_{-3.66}^{+3.35}$ & $121.79_{-4.36}^{+4.84}$ \\
    & 2629 & $86.86_{-29.49}^{+28.63}$ & $84.18_{-33.94}^{+34.90}$ & $5.23_{-0.98}^{+0.53}$ & $5.23_{-1.26}^{+1.01}$ & $97.27_{-0.34}^{+0.33}$ & $98.01_{-0.45}^{+0.44}$ & $125.96_{-4.03}^{+4.39}$ & $125.04_{-5.81}^{+6.25}$ \\
    & 3111 & $114.30_{-26.66}^{+25.23}$ & $96.70_{-31.69}^{+29.33}$ & $5.06_{-1.60}^{+0.74}$ & $4.97_{-1.08}^{+0.71}$ & $97.24_{-0.31}^{+0.23}$ & $97.97_{-0.35}^{+0.30}$ & $128.16_{-3.80}^{+3.13}$ & $128.37_{-4.46}^{+4.46}$ \\
    & 3725 & $99.66_{-36.43}^{+31.96}$ & $89.09_{-41.37}^{+40.69}$ & $3.59_{-1.04}^{+0.63}$ & $7.09_{-2.37}^{+2.30}$ & $96.97_{-0.27}^{+0.22}$ & $97.20_{-2.27}^{+1.27}$ & $135.15_{-5.00}^{+4.51}$ & $111.42_{-14.69}^{+13.20}$ \\
    \hline
     \multirow{8}{*}{58738}
     & 879 & $101.46_{-19.24}^{+18.95}$ & $92.35_{-26.20}^{+21.48}$ & $6.19_{-0.94}^{+0.45}$ & $6.97_{-0.84}^{+0.59}$ & $265.57_{-0.28}^{+0.26}$ & $266.18_{-0.47}^{+0.37}$ & $130.01_{-3.17}^{+3.19}$ & $126.57_{-4.25}^{+4.06}$ \\
     & 1229 & $126.63_{-22.49}^{+22.18}$ & $107.65_{-22.56}^{+23.49}$ & $4.15_{-1.57}^{+1.16}$ & $5.16_{-1.30}^{+0.60}$ & $265.32_{-0.22}^{+0.15}$ & $266.07_{-0.26}^{+0.21}$ & $135.29_{-3.33}^{+2.36}$ & $133.81_{-3.79}^{+3.63}$ \\
     & 1579 & $113.99_{-21.50}^{+18.50}$ & $111.59_{-24.99}^{+19.75}$ & $5.19_{-1.14}^{+0.69}$ & $5.16_{-1.21}^{+0.69}$ & $264.69_{-0.25}^{+0.19}$ & $265.60_{-0.29}^{+0.20}$ & $129.04_{-3.01}^{+2.48}$ & $130.65_{-3.77}^{+3.27}$ \\
     & 1972 & $102.99_{-24.51}^{+21.64}$ & $90.82_{-24.78}^{+25.46}$ & $5.35_{-0.98}^{+0.51}$ & $5.83_{-0.93}^{+0.74}$ & $264.75_{-0.30}^{+0.22}$ & $265.78_{-0.35}^{+0.33}$ & $130.69_{-3.91}^{+3.40}$ & $128.92_{-4.46}^{+4.69}$ \\
     & 2279 & $112.87_{-28.56}^{+26.86}$ & $84.69_{-31.44}^{+30.89}$ & $6.49_{-2.09}^{+0.93}$ & $6.05_{-1.32}^{+1.05}$ & $263.95_{-0.49}^{+0.39}$ & $265.12_{-0.48}^{+0.48}$ & $120.57_{-4.37}^{+4.25}$ & $123.88_{-5.65}^{+5.70}$ \\
     & 2629 & $88.71_{-35.01}^{+36.77}$ & $90.47_{-32.38}^{+35.22}$ & $5.55_{-1.34}^{+0.84}$ & $5.14_{-1.32}^{+1.16}$ & $263.73_{-0.47}^{+0.45}$ & $264.71_{-0.42}^{+0.41}$ & $123.49_{-4.96}^{+5.50}$ & $126.10_{-6.16}^{+6.52}$ \\
     & 3111 & $83.99_{-28.58}^{+27.78}$ & $93.92_{-38.59}^{+38.12}$ & $4.60_{-0.78}^{+0.40}$ & $4.69_{-1.36}^{+1.07}$ & $263.33_{-0.25}^{+0.26}$ & $264.71_{-0.40}^{+0.38}$ & $127.57_{-3.26}^{+3.36}$ & $128.02_{-6.21}^{+6.45}$ \\
     & 3725 & $78.61_{-37.21}^{+38.03}$ & $87.93_{-39.80}^{+40.98}$ & $4.77_{-1.54}^{+1.17}$ & $3.16_{-1.18}^{+1.44}$ & $263.62_{-0.41}^{+0.44}$ & $264.75_{-0.37}^{+0.50}$ & $125.69_{-6.03}^{+6.49}$ & $136.12_{-8.55}^{+7.36}$ \\
     \hline
    \end{tabular}
\end{table*}

Interestingly, the intersection point of the power-law curves for the states A and B changed significantly between the two observations. In the first observation, the two power-law curves intersected at approximately 514 MHz, which is consistent with the 500 MHz reported by \citep{2020ApJ...890...31Z}, but in the second observation, the intersection frequency changed to approximately 1044 MHz. The frequency corresponding to this intersection point can be regarded as a transition frequency, indicating that state A becomes brighter than state B.

As shown in Figure \ref{cyq8}, the relative brightness between the two states at the center frequency of 879 MHz differs between the two observations. Specifically, state A is slightly brighter than state B in the first observation, whereas state B is brighter in the second observation. To investigate this phenomenon, we integrated the single pulse in two emission states, and the integral profile is shown in Figure \ref{cyq9}. The left and right columns show the results of the first and second observations, respectively, with a center frequency of 879 MHz. From the top panel of Figure \ref{cyq9}, it can be seen that in the first observation, the peak flow density of state A was slightly higher than that of state B, while the opposite was true in the second observation. This phenomenon is consistent with \citet{2016MNRAS.462.2518R} but with different frequencies from their paper. Please note that there may be biases when comparing the peak intensity of the profiles between the two observations due to factors such as the signal-to-noise ratio or the total number of sub-integrations between the two states. However, in general, the above phenomenon strongly suggests that the relative brightness between the two emission states of PSR J0614+2229 is time-dependent, which is difficult to understand using existing models.

\begin{figure*}[h!]
    \centering
    \includegraphics[width=7in,height=9in]{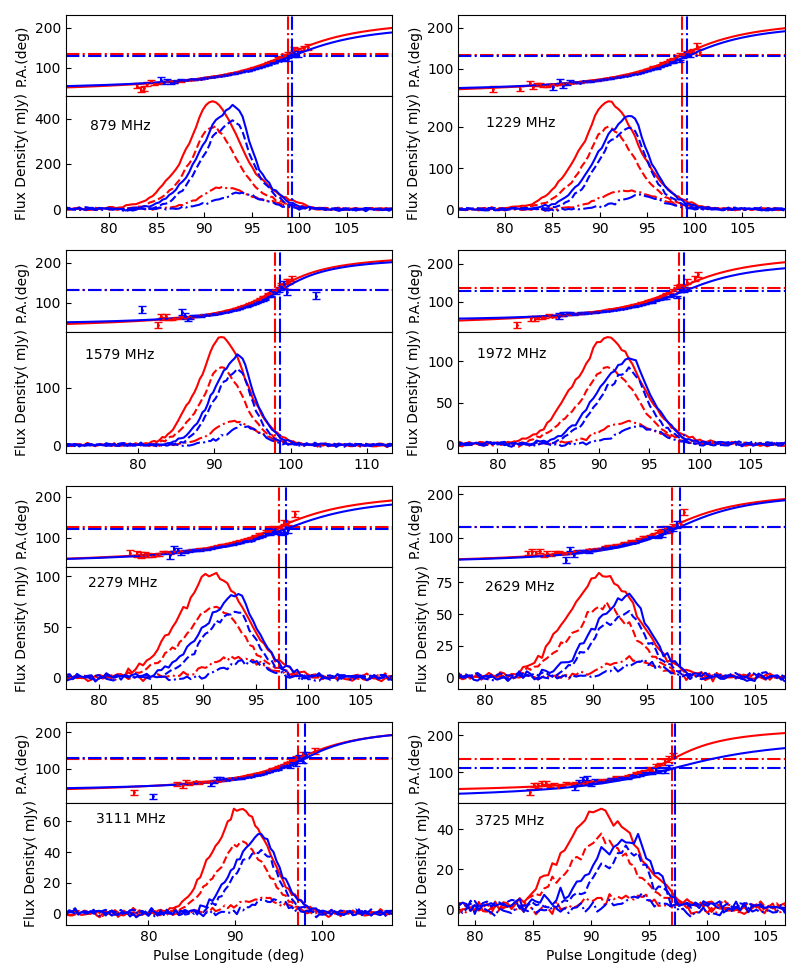}
    \caption{The bottom panels represent the averaged polarization profiles of state A (red lines) and state B (blue lines) for the first observation at eight frequencies. Where the total power, the total linear polarization, and the circular polarization are represented by the solid line, dashed line, and dash-dotted line, respectively. The red and blue dots, in the upper panels, correspond to the linear polarization angles for state A and state B alone with the best-fit curve from the RVM fit shown as the red and blue curves. The red and blue vertical dash-dotted lines show the steepest gradient point phase of the RVM fit, $\phi_0$, for states A and B.}
    \label{cyq10}
\end{figure*}

\begin{figure*}[h!]
    \centering
    \includegraphics[width=7in,height=9in]{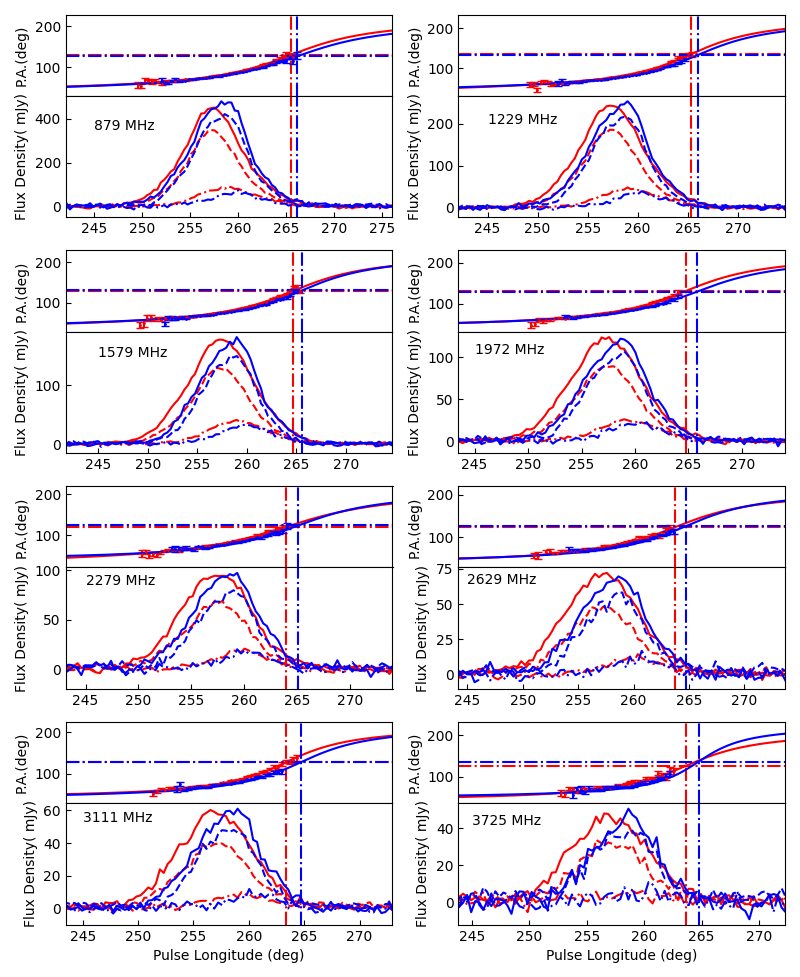}
    \caption{Same as Figure \ref{cyq10} but for the second observation}
    \label{cyq11}
\end{figure*}

\subsection{Polarisation} \label{sec5}
\begin{figure}
    \centering
    \includegraphics[width=\columnwidth]{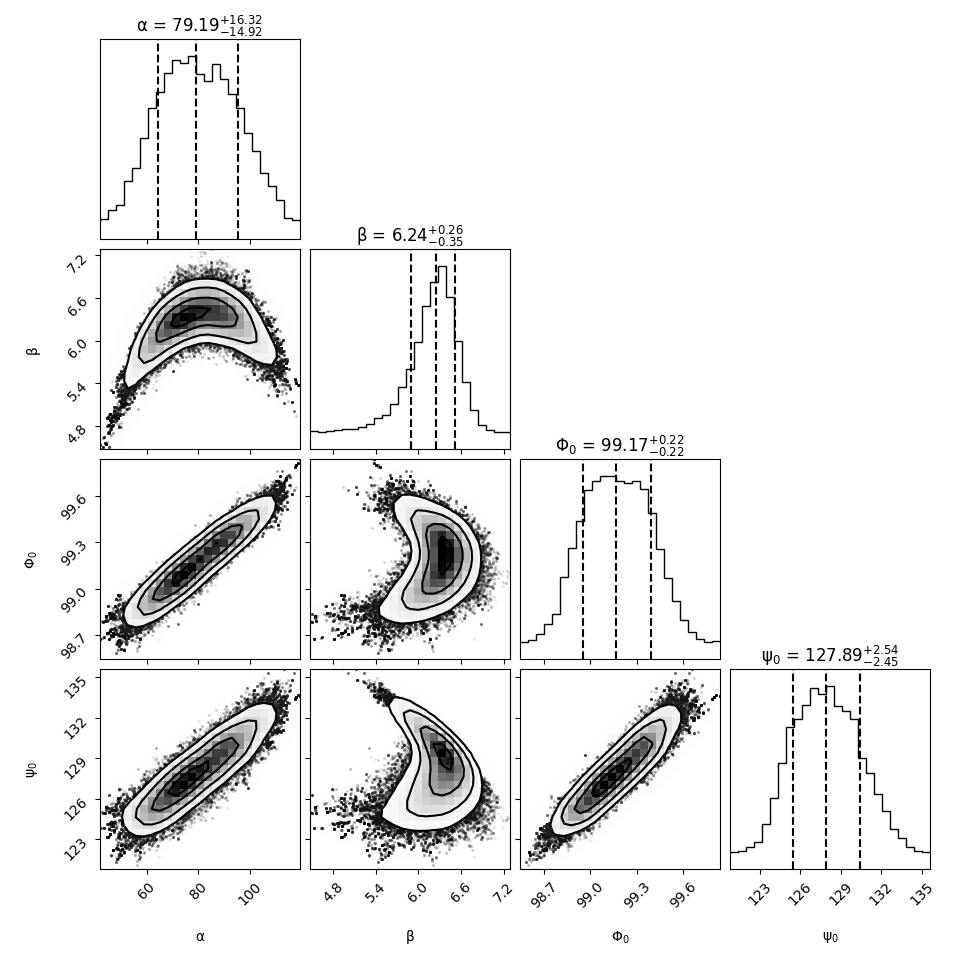}
    \caption{Posterior distributions of the four parameters included in RVM fits as resulting from MCMC for state B at the center frequency on 879 MHz in the first observation. The one-dimensional posterior for each parameter is shown at the top of each column, and the other plots represent the corresponding correlations between these four parameters. The dashed vertical lines indicate the distribution's median, as well as its 16th and 84th percentiles, respectively. All angles' units are degrees.}
    \label{cyq12}
\end{figure}

We obtained the polarization profiles of the states A and B, and the corresponding linear polarization position angles, as shown in Figures \ref{cyq10} (for the first observation) and \ref{cyq11} (for the second observation). The bottom panel shows the average polarization profiles of states A and B. The red and blue solid, dashed, and dotted lines represent the total intensity (I), total linear polarization ($L=\sqrt{Q^{2}+U^{2}}$), and circular polarization ($V$) of states A and B, respectively. It is evident that the pulsar exhibits a highly linear polarization fraction across a wide range of frequencies, and the circular polarization fraction decreases with the increase in frequency. The red and blue dots in the top panel represent the linear polarization position angles $({\rm \Psi} =\frac{1}{2}\arctan(\frac{U}{Q})$) of the A and B modes, respectively. To obtain the information of the viewing geometry, we fit the linear polarization angles for states A and B with the Rotating Vector Model \citep[RVM;][]{1969ApL.....3..225R} as follows: 
\begin{equation}
\tan(\psi - \psi_0) = \frac{\sin\alpha \sin(\phi - \phi_0)}{\sin(\xi) \cos\alpha - \cos(\xi) \sin\alpha \cos(\phi - \phi_0)},
\end{equation}
where $\xi$ = $\alpha$ + $\beta$, $\alpha$ is the angle between the rotation axis and the magnetic axis, and $\beta$ is the angle between the rotation axis and the observer's line of sight. $\psi$ is the polarization position angle at a pulse longitude $\phi$, $\psi_0$ and $\phi_0$ are the phase offsets for polarization position angle and pulse rotational phase, respectively. Here, we perform Markov Chain Monte Carlo fitting (\citeauthor{2019MNRAS.490.4565J} \citeyear{2019MNRAS.490.4565J}) using the python package \textsc{emcee} (\citeauthor{2013PASP..125..306F} \citeyear{2013PASP..125..306F}) to determine $\alpha$, $\beta$, $\phi_0$ and $\psi_0$. An example `corner' plot of the post distribution for state B at the center frequency of 879 MHz in the first observation is shown in Figure \ref{cyq12}. Results of the RVM fitting at the 16 and 84 percentiles of the posterior probability distributions are given in Table \ref{tablefour}, where the $\phi_0$ and $\psi_0$ can be excellently constrained. However, due to the limited duty cycle of the pulse profile (\citeauthor{2004A&A...421..215M} \citeyear{2004A&A...421..215M}), it posed considerable challenges in determining the geometrical angles as $\alpha$ and $\beta$ exhibited significant covariance and unreliability. The best fits of the RVM are shown as red (state A) and blue (state B) curves in the top panels of Figure \ref{cyq10} and Figure \ref{cyq11}, which both show a better S-shape and PPA swings are different. There is a way developed by \citet{1991ApJ...370..643B} to estimate the emission heights from the center of the star with significant advantages. The absolute emission height at the frequency of observation could be converted from the net delay predicted by aberration and retardation (A/R, \citeauthor{2004ApJ...607..939D} \citeyear{2004ApJ...607..939D}; \citeauthor{2015MNRAS.446.3356R} \citeyear{2015MNRAS.446.3356R}), and is given by 
\begin{equation}
h_{em} = \frac{Pc\Delta\phi}{8\pi},
\end{equation}
where $P$ is the rotation period of the pulsar and $c$ is the speed of light. $\Delta\phi$ = $\phi_0$ $-$ $\phi_{fid}$, the estimate of $\phi_{fid}$ can be obtained based on the pulse profile morphology. The derived emission heights for states A and B are given in Table \ref{tablefive}. It is evident that the emission height of state A in both observations is higher than that of state B.

\begin{table*}
\renewcommand{\arraystretch}{1.5}
   \centering
    \caption{Emission height and the half opening angle estimate for state A and state B}
    \label{tablefive}
    \begin{tabular}{cccccccccc}
    \hline
    \hline
    MJD & State & $r_{879}$ & $r_{1229}$ & $r_{1579}$ & $r_{1972}$ & $r_{2279}$ & $r_{2629}$ & $r_{3111}$ & $r_{3725}$ \\
    (d) &  & (km) & (km) & (km) & (km) & (km) & (km) & (km) & (km) \\
    \hline
     \multirow{2}{*}{58710}
     & A & $543.30_{-18.13}^{+11.85}$ & $527.61_{-14.64}^{+11.16}$ &  $484.34_{-12.55}^{+10.46}$ &$484.22_{-12.55}^{+6.97}$ &$480.23_{-21.62}^{+18.83}$ &$457.67_{-23.71}^{+23.01}$ &$456.08_{-16.04}^{+16.04}$ &$412.72_{-15.34}^{+15.34}$ \\
     & B &$425.72_{-16.74}^{+13.95}$ &$416.47_{-15.34}^{+14.64}$&$383.51_{-11.85}^{+11.85}$& $370.93_{-13.95}^{+16.74}$& $333.32_{-27.19}^{+27.89}$& $337.59_{-31.38}^{+30.68}$& $359.58_{-24.41}^{+20.92}$& $257.22_{-158.29}^{+88.56}$\\
     \hline
     \multirow{2}{*}{58738}
    & A & $524.75_{-19.53}^{+18.13}$ & $507.28_{-15.34}^{+10.46}$ & $463.76_{-17.43}^{+13.25}$ &$491.73_{-20.92}^{+15.34}$ & $411.87_{-34.17}^{+27.19}$ &$420.84_{-32.78}^{+31.38}$ & $417.59_{-17.43}^{+18.13}$ &$413.34_{-28.59}^{+30.68}$ \\
    &B & $469.39_{-32.78}^{+25.80}$& $461.39_{-18.13}^{+14.64}$& $428.74_{-20.22}^{+1395}$& $465.56_{-24.41}^{+23.01}$& $395.31_{-33.47}^{+33.47}$ & $366.68_{-29.29}^{+28.59}$& $366.44_{-27.89}^{+26.50}$ & $394.21_{-25.80}^{+34.85}$ \\
    \hline
    \end{tabular}
\end{table*}

\section{Discussion} \label{secfour}
\subsection{Quasi-stable magnetosphere?}
As shown in Figure \ref{cyq4} and Figure \ref{cyq5}, the time-dependent mode switching for PSR J0614+2229 presents a gradual shift pattern of the emission longitude. This phenomenon is very similar to the 'swooshes' events in PSRs B1859+07 and B0919+06 (\citeauthor{2006MNRAS.370..673R} \citeyear{2006MNRAS.370..673R}), which have been studied in great detail over the last decade (\citeauthor{2016MNRAS.461.3740W} \citeyear{2016MNRAS.461.3740W}; \citeauthor{2016MNRAS.456.3413H} \citeyear{ 2016MNRAS.456.3413H}; \citeauthor{2017MNRAS.469.2049Y} \citeyear{2017MNRAS.469.2049Y}). \citeauthor{2021MNRAS.506.5836R} (\citeyear{2021MNRAS.506.5836R}) described a simple emission model to explain most of the key emission features of PSRs B1859+07 and B0919+06 by shrinking and expanding the magnetosphere. They proposed that their emission model can be invoked to explain the fast mode changes in PSR J0614+2229, namely the origin of the mode switching for this pulsar may be related to periodic shrinkage or expansion of the magnetosphere. A similar mechanism for this pulsar is also mentioned in a previous paper (\citeauthor{2016MNRAS.462.2518R} \citeyear{2016MNRAS.462.2518R}) but not answered with certainty. Polarimetric observations may provide new information to understand the underlying physics of this pulsar’s emission process. 
By fitting PAA with RVM, we obtained the values of $\alpha$ and $\beta$ for the two emission states of this pulsar.
These values are shown in columns $3-6$ in Table \ref{tablefour}, where the subscripts "A" and "B" represent state A and state B, respectively. From Table \ref{tablefour}, the $\alpha$ and $\beta$ values for the two emission states have not changed significantly within the uncertainty in the two observations. 
This implies that the geometry of the emission region for the two emission states may not have changed.
Furthermore, switching from state A to state B only requires about a few pulse periods in this pulsar. It is difficult to understand the magnetosphere geometry of the pulsar switches between two different states in such short timescales. Therefore，we speculate that the geometry of the radio emission region in the pulsar magnetosphere does not change when PSR J0614+2229 undergoes a mode switching, that is, the mode switching of this pulsar is unlikely to be related to changes in the magnetosphere. Of course， another situation is that perhaps the changes in the magnetosphere are too small to make the geometry parameters in the radio emission region change. \citeauthor{2010Sci...329..408L} (\citeyear{2010Sci...329..408L}) presented a direct connection between the radio emission variability and the spin-down rate for six pulsars suggesting the changing currents of charged particles in the pulsar magnetosphere that alters the emission mechanism. Measuring the spin-down rate corresponding to the two emission states of PSR J0614+2229 may provide some clues to understanding the underlying physics of this pulsar's emission process. However, this work is beyond the scope of this paper due to the short observation time. 
Therefore, it is necessary to conduct long-term pulsar timing monitoring on this pulsar to study whether its spin-down rate change is associated with the switching between the two states.

\subsection{Probing the mode switching mechanism}
\begin{figure}
    \centering
    \includegraphics[width=\columnwidth]{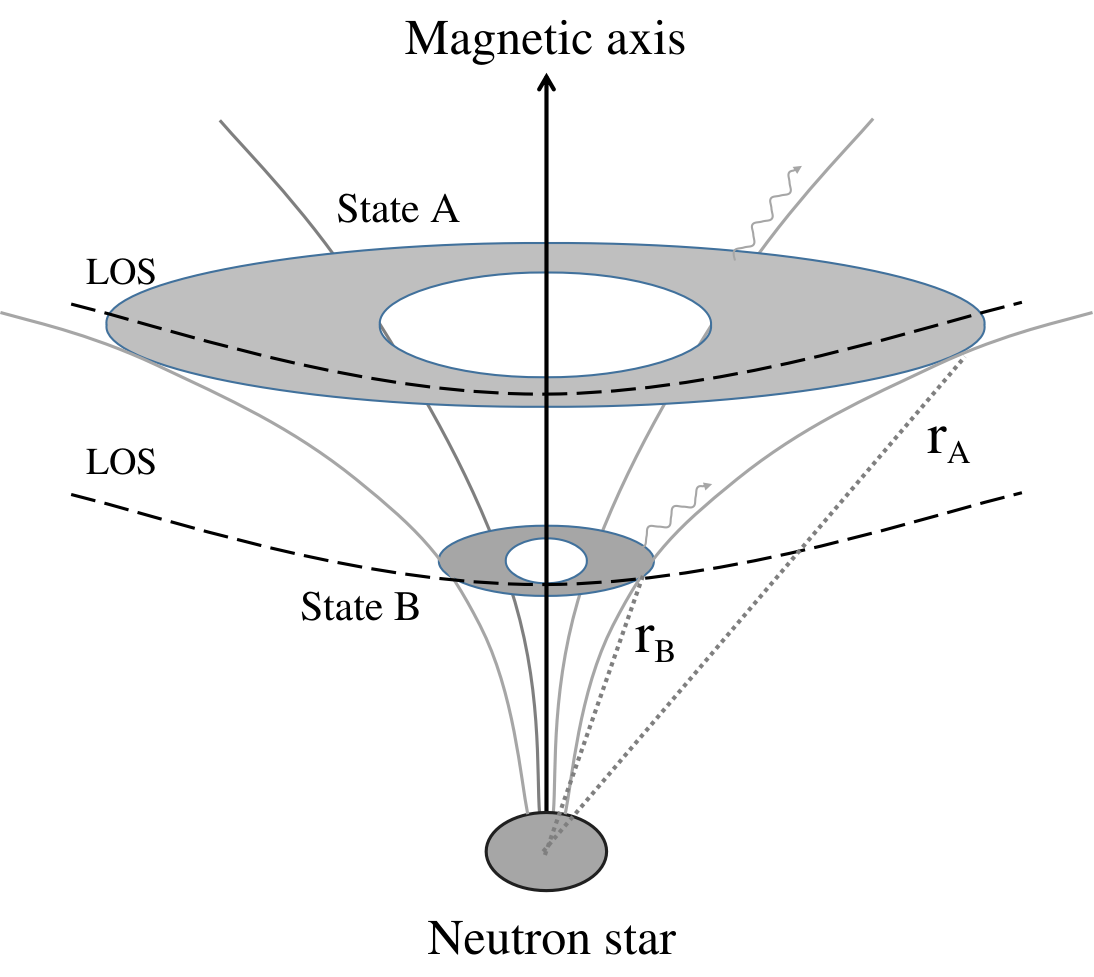}
    \caption{Schematic diagram explaining the observed behavior of two emission states of PSR J0614+2229. The state A and state B emission regions, symmetrical around the magnetic axis, are assumed to be generated at different altitudes, $r_{A}$ and $r_{B}$, respectively. The beam size expands during state A, as shown by the trajectories of the line of sight (LOS).}
    \label{cyq13}
\end{figure}

The emission properties of PSR J0614+2229 exhibit great complexity. The most significant feature is that the mode change in this pulsar is much faster than other pulsars. Based on the highly sensitive observation using the Five-hundred-meter Aperture Spherical radio Telescope, \citeauthor{2022ApJ...934...57S} (\citeyear{2022ApJ...934...57S}) attributed the mode switching of this pulsar to the changes in emission heights. Similar to another mode switching pulsar PSR J1326$-$6700, which shows a gradual earlier shift pattern of the emission longitude and \citeauthor{2020ApJ...904...72W} (\citeyear{2020ApJ...904...72W}) proposed that the abnormal mode of this pulsar might originate from higher altitudes than the normal mode. For PSR J0614$+$2229， the delay emission heights for states A and B derived using the relativistic beaming model based on the A/R effect are given in Table \ref{tablefive}. It is found that state A emission arises at a higher altitude from the surface of the neutron star while state B arises closer to the stellar surface at a ﬁxed frequency. Here, a possible model is constructed, wherein the emission of both states comes from the same magnetic ﬂux surface, but from different heights at a ﬁxed frequency \citep{2003A&A...399..223V,2020ApJ...904...72W}. As shown in Figure \ref{cyq13}, the emission height at a ﬁxed frequency increases when the emission changes from state B to state A. Over a wide frequency range, the emission height in both states did not significantly change within the uncertainty range. This phenomenon supports a broadband interpretation in which the broadband emission is assumed to come from a narrow range of altitude \citep{2010MNRAS.401.1781D}, which implies that the relativistic charged particles of this pulsar ﬂowing out along a magnetic ﬂux tube may generate wide-band radio emission. 

\begin{figure}
    \centering
    \includegraphics[width=\columnwidth]{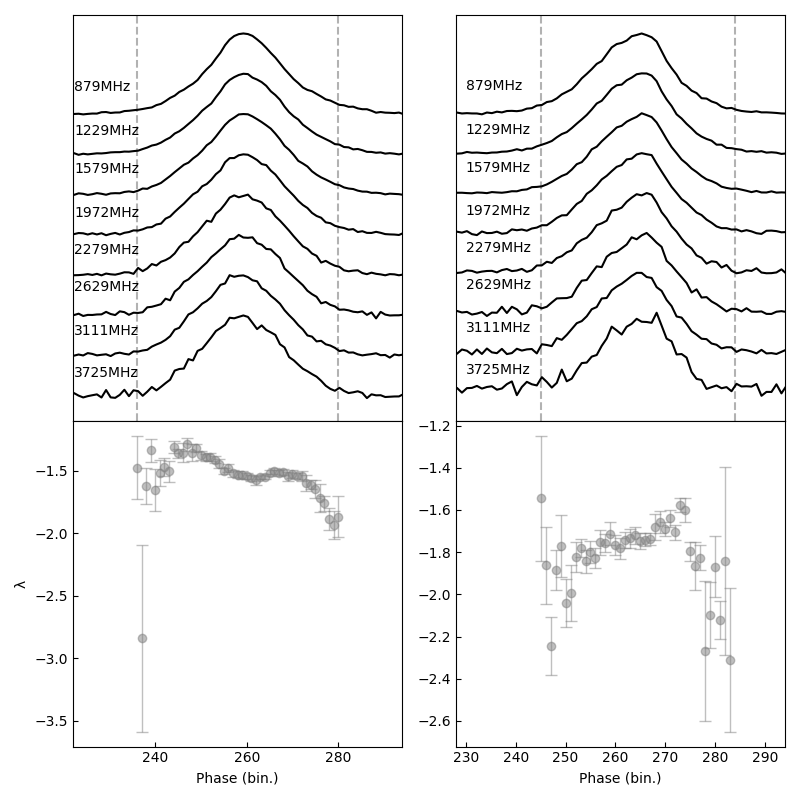}
    \caption{The phase-resolved spectrum of the first observation. The upper panels are phase-aligned multi-frequency profiles of states A (left panel) and B (right panel), respectively. The lower panels are corresponding phase-resolved spectra derived from the phase-aligned multi-frequency profiles. The dashed lines in the upper panels represent the longitude boundaries of the phase-resolved spectrum.}
    \label{cyq14}
\end{figure}

\begin{figure}
    \centering
    \includegraphics[width=\columnwidth]{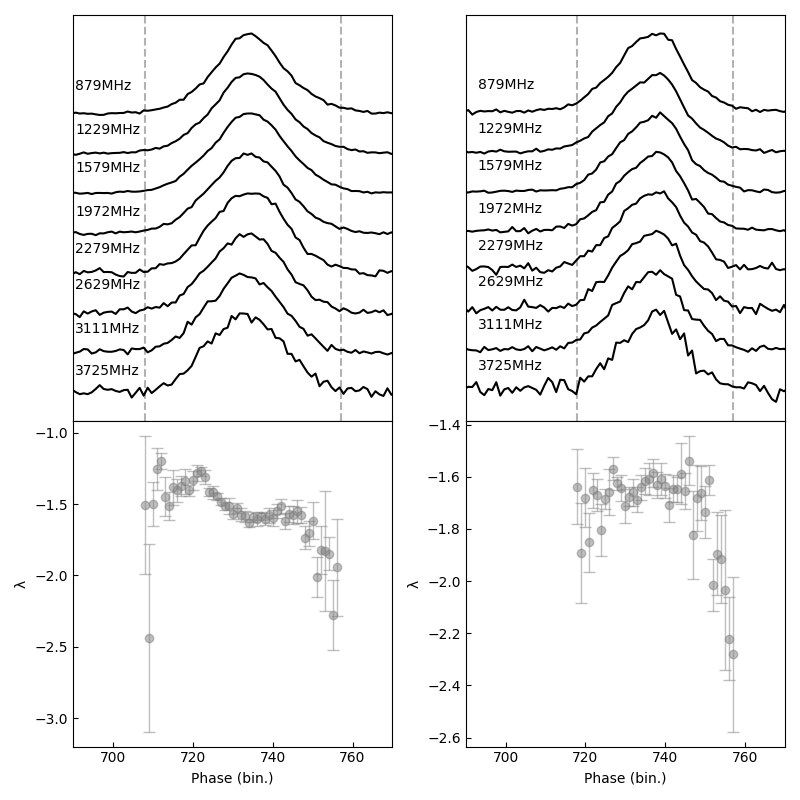}
    \caption{Same as Figure \ref{cyq14} but for the second observation.}
    \label{cyq15}
\end{figure}

The difference in the frequency-dependent behaviors of intensity and pulse width between state A and state B may imply the distributions of the spectral index of the phase-resolved spectra (PHRS) corresponding to state A and state B are different. 
We employ the method from \citet{2007ChJAA...7..789C} and obtained spectral index. The detailed processes are as follows: (1) In the emission windows of states A and B, for the appointed phase bin: the two states' relative intensity ratio at each frequency is defined as $\eta_f$ = $I_f$/$I_{1579}$, where $I$ is the intensity, and the subscripts "f" and "1579" denoting the observing and reference frequencies, respectively. With the function of log$\eta_f$ = $\lambda$log$f$ + C, the relative spectral index $\lambda$ can be obtained by least square fit to the data of $\eta_f$ and $f$. (2) Calculate $\lambda$ for each bin in the pulse emission windows of states A and B through step (1). The phase-resolved spectra from the first and second observations are plotted in the lower panels of Figure \ref{cyq14} and Figure \ref{cyq15}, where the left and right columns correspond to the results of state A and state B, respectively. The upper panels of these two figures represent the phase-aligned multi-frequency profiles and the dotted vertical lines show the longitude boundaries of the phase-resolved spectra. To further examine the PHRS, we averaged the $\lambda$ we obtained for states A and B. For the first observation, the mean values of $\lambda$ for states A and B are $-$1.55 and $-$1.82, which are well consistent with the best-fitting spectral indices $-$1.53(2) and $-$1.83(5) in Figure \ref{cyq8}. For the second observation, the mean values of $\lambda$ for states A and B are $-$1.57 and $-$1.72, respectively, which are also well consistent with $-$1.55(3) and $-$1.78(5) in Figure \ref{cyq8}. The shapes of the phase-resolved spectra for state A in the two observations show that the spectrum steepens from the edge to the center of the emission region if we ignore the $\lambda$ with a larger error at the edge, this just explains the pulse width of state A generally increases with frequency (\citeauthor{2020ApJ...890...31Z} \citeyear{2020ApJ...890...31Z}; \citeauthor{2014ApJS..215...11C} \citeyear{2014ApJS..215...11C}). However, for state B in the two observations, the longitudinal spectra seem relatively stable from the center to the edge of the emission region, which may be the reason why the $W_{50}$ of this state versus frequency is also relatively stable at a wide frequency range from 0.7 GHz to 4 GHz. In general, there is a significant difference in the distributions of spectral index across the emission regions of states A and B. Therefore, we speculate that the mode switching of PSR J0614+2229 is possibly caused by changes in the emission heights that alter the distributions of spectral index across the emission regions of states A and B resulting in the frequency-dependent behaviors, i.e., intensity and pulse width.

\section{conclusions} \label{secfive}
We have carried out a detailed analysis on PSR J0614+2229 based on the data observed with the UWL receiver on the Parkes radio telescope. The multi-frequency data reveal a wealth of information about the emission characteristics of this pulsar. Same as previous observations, the $W_{50}$ of state B decreases with increasing frequency, while that of state A is the opposite, and both states follow a simple power law very well. With further research, we found that the variation of the relative brightness between the two states is time-related. 
The total linear polarization, $\alpha$, $\beta$, and the PPA of two emission states do not significantly change within the uncertainty range at a wide frequency range. This implies that the geometry of the radiation region may not have changed when this pulsar undergoes a mode switching. {The emission height of state A is estimated to be higher than that of state B at a fixed frequency.} Using the phase-aligned multi-frequency proﬁles, we estimated the PHRS spectral index across the pulse phase and found that the types of spectral-index variation across the emission regions (in the longitudinal direction) for states A and B are different. Therefore, we speculate that the mode switching of this pulsar is possibly caused by changes in the emission heights that alter the distributions of spectral index across the emission regions of states A and B resulting in the frequency-dependent behaviors, i.e., intensity and pulse width. 
We expect to conduct long-term and frequent pulsar timing observations of this pulsar to investigate the correlation between its spin-down rate switch and emission states switch. These observations will provide crucial observational evidence for studying the physical mechanisms underlying the switching of its emission states.


\section{Acknowledgments}

This work is supported by the Major Science and Technology Program of Xinjiang Uygur Autonomous Region (No.2022A03013-4,2022A03013-2), the Guizhou Province Science and Technology Foundation (Nos. ZK[2022]304,ZK[2023]024), the Scientific Research Project of the Guizhou Provincial Education (Nos. KY[2022]132, KY[2022]137), the National Natural Science Foundation (Grant Nos.: 12273008, U1731238, U1831120, U1838106, 61875087, 61803373, 11303069, 11373011, 11873080, 12103069), the Foundation of Guizhou Provincial Education Department (No. KY (2020)003), the National Key R\&D Program of China (No. 2017 YFB 0503300), the Foundation of Science and Technology of Guizhou Province (No. [2016]4008, [2017]5726-37, [2018]5769-02), and the Foundation of Guizhou Provincial Education Department (No. KY(2020)003). The Parkes radio telescope is part of the Australia Telescope National Facility which is funded by the Australian Government for operation as a National Facility managed by CSIRO. This paper includes archived data obtained through the Australia Telescope Online Archive and the CSIRO Data Access Portal (\url{http://data.csiro.au}).


\end{CJK}
\bibliography{sample631}{}
\bibliographystyle{aasjournal}

\end{document}